\documentclass[amsmath,aps,showpacs,a4paper,10pt]{revtex4}

 \usepackage{epsf}
 \usepackage{graphicx}    

 \usepackage{enumerate}   

\begin{document}

 \newcommand{\be}[1]{\begin{equation}\label{#1}}
 \newcommand{\ee}{\end{equation}}
 \newcommand{\bea}{\begin{eqnarray}}
 \newcommand{\eea}{\end{eqnarray}}
 \def\disp{\displaystyle}

 \def\gsim{ \lower .75ex \hbox{$\sim$} \llap{\raise .27ex \hbox{$>$}} }
 \def\lsim{ \lower .75ex \hbox{$\sim$} \llap{\raise .27ex \hbox{$<$}} }

 \begin{titlepage}

 \begin{flushright}
 arXiv:1806.02773
 \end{flushright}

 \title{\Large \bf Null Signal for the Cosmic Anisotropy in
 the~Pantheon~Supernovae Data}

 \author{Hua-Kai~Deng\,}
 \email[\,email address:\ ]{dhklook@163.com}
 \affiliation{School of Physics,
 Beijing Institute of Technology, Beijing 100081, China}

 \author{Hao~Wei\,}
 \email[\,Corresponding author;\ email address:\ ]{haowei@bit.edu.cn}
 \affiliation{School of Physics,
 Beijing Institute of Technology, Beijing 100081, China}

 \begin{abstract}\vspace{1cm}
 \centerline{\bf ABSTRACT}\vspace{2mm}
 The cosmological principle assumes that the universe is homogeneous
 and isotropic on cosmic scales. There exist many works testing the
 cosmic homogeneity and/or the cosmic isotropy of the universe
 in the literature. In fact, some observational hints of the cosmic
 anisotropy have been claimed. However, we note that the paucity of
 the data considered in the literature might be responsible for the
 ``found'' cosmic anisotropy. So, it might disappear in a large
 enough sample. Very recently, the Pantheon sample consisting
 of 1048 type Ia supernovae (SNIa) has been released, which is
 the largest spectroscopically confirmed SNIa sample to date.
 In the present work, we test the cosmic anisotropy in the Pantheon
 SNIa sample by using three methods, and hence the results from
 different methods can be cross-checked. All the results obtained by
 using the hemisphere comparison (HC) method, the dipole fitting (DF)
 method and HEALPix suggest that no evidence for the cosmic
 anisotropy is found in the Pantheon SNIa sample.
 \end{abstract}


 \maketitle

 \end{titlepage}

 \renewcommand{\baselinestretch}{1.0}


\section{Introduction}\label{sec1}

In modern cosmology, it is usually assumed that the universe
 is homogeneous and isotropic on cosmic scales~\cite{Weinberg,
 Kolb90}. This is the well-known cosmological principle, which
 plays a fundamental role. Although it is indeed a very
 good approximation across a vast part of the universe (see
 e.g.~\cite{Hogg:2004vw,Hajian:2006ud}), the
 cosmological principle has not yet been well proven on the
 scales $\gsim\,1\,$Gpc~\cite{Caldwell:2007yu}. Therefore, it
 is still of interest to carefully test both the homogeneity
 and the isotropy of the universe.

The cosmic homogeneity can be broken in the well-known
 Lema\^{\i}tre-Tolman-Bondi~(LTB) void model~\cite{LTB} and
 others akin to it. In such kind of models, the cosmic acceleration
 could be explained without the needs of dark energy or modification
 to general relativity. Thus, the LTB-like models have been
 extensively studied in the literature. On the other hand, actually
 the cosmic homogeneity has been tested by using various
 observations. For conciseness, we refer to
 e.g.~\cite{Yan:2014eca,Deng:2018yhb} and
 references therein for details.

In the present work, we are mainly interested in
 the cosmic isotropy. It can be broken in the models of
 G\"odel universe~\cite{Godel:1949ga}, Bianchi type I $\sim$ IX
 universes~\cite{Bianchi}, Finsler universe~\cite{Li:2015uda},
 and so on. In fact, some observational hints of the cosmic
 anisotropy have been claimed in the literature. For instance,
 it has been found that there is a preferred axis in the cosmic
 microwave background (CMB) temperature map (known as the
 ``Axis of Evil'' in the literature)~\cite{Axisofevil,
 Zhao:2016fas,Hansen:2004vq,Campanelli:2006vb,
 Ade:2013nlj,Ade:2015hxq,Schwarz:2015cma}. Another kind of
 hints for the cosmic anisotropy comes from the distribution of type
 Ia supernovae (SNIa)~\cite{Schwarz:2007wf,Antoniou:2010gw,
 Mariano:2012wx,Cai:2011xs,Zhao:2013yaa,Yang:2013gea,Chang:2014nca,
 Lin:2015rza,Lin:2016jqp,Chang:2017bbi,Javanmardi:2015sfa,
 Bengaly:2015dza,Deng:2018yhb,Wang:2017ezt,Sun:2018epo,
 Heneka:2013hka}. It is found that there exists a preferred
 direction (or more) in various SNIa datasets (e.g. Union2,
 Union2.1, JLA), mainly by using the hemisphere comparison (HC)
 method proposed in~\cite{Schwarz:2007wf} and then improved
 by~\cite{Antoniou:2010gw}, as well as the dipole fitting (DF)
 method proposed in~\cite{Mariano:2012wx}. Also, these works
 have been extended to include gamma-ray bursts
 (GRBs)~\cite{Meszaros:2009ux,Wang:2014vqa,Chang:2014jza}, and
 the preferred direction still exists. On the other hand, it is
 claimed in 1998 that the fine structure ``constant'' $\alpha$
 is time-varying~\cite{Webb98,Webb00} (see also
 e.g.~\cite{{Uzan10,Barrow09,HWalpha,Wei:2016moy}}). A dozen of
 years later, a preferred direction was also found in the
 $\Delta\alpha/\alpha$ data~\cite{King:2012id,Webb:2010hc},
 which means that the fine structure ``constant'' $\alpha$ is
 also spatially varying. Note in~\cite{Mariano:2012wx} that the
 preferred direction in the $\Delta\alpha/\alpha$ data might
 be correlated with the one in the distribution of SNIa.
 Furthermore, similar preferred direction(s) has/have also been
 found in other observational data, such as rotationally
 supported galaxies~\cite{Zhou:2017lwy,Chang:2018vxs}, quasars and
 radio galaxies~\cite{Singal:2013aga,Bengaly:2017slg}, as well
 as the quasar optical polarization data~\cite{Hutsemekers,
 Pelgrims:2016mhx}. Actually, in Table~I and
 Fig.~10 of~\cite{Deng:2018yhb}, we summarized the preferred
 directions in various observational data mentioned above. Most
 of them are located in a relatively small part (about a
 quarter) of the north galactic hemisphere (see~\cite{Deng:2018yhb}
 for details). In some sense, they are in agreement with each
 other, and hence this fact suggests that the possible
 cosmic anisotropy should be taken seriously.

However, it is worth noting that the paucity of the data
 mentioned above might be responsible for the ``found'' cosmic
 anisotropy. For example, the numbers of SNIa in
 the Union2~\cite{Amanullah:2010vv}, Union2.1~\cite{Suzuki:2011hu} and
 JLA~\cite{Betoule:2014frx} datasets are 557, 580 and 740,
 respectively. The number of usable GRBs (without
 the circularity problem) is only 59~\cite{Wei:2010wu,
 Chang:2014jza}, 79~\cite{Liu:2014vda}, or 116~\cite{Wang:2014vqa}.
 The number of usable $\Delta\alpha/\alpha$ data~\cite{King:2012id,
 Wei:2016moy} is 293. The number of SPARC rotationally
 supported galaxies~\cite{Lelli:2016zqa} is 175. Obviously, the
 numbers of the data points mentioned above are $\sim{\cal O}(10^2)$,
 which are not enough to form statistically large samples. It
 is reasonable to imagine that the ``found'' cosmic anisotropy
 might be just caused by statistical fluctuations, and might
 disappear in a large enough sample.

Very recently, the Pantheon sample~\cite{Scolnic:2017caz,
 Pantheondata,Pantheonplugin} consisting of 1048 SNIa has been
 released, which is the largest spectroscopically confirmed
 SNIa sample to date. In fact, it is the first
 sample containing $\sim{\cal O}(10^3)$ SNIa. Therefore, it
 is very interesting to test the cosmic anisotropy in the
 Pantheon SNIa sample, with the hope to find something
 different.

In this work, we test the cosmic anisotropy in the Pantheon
 SNIa sample by using three methods, and hence the results from
 different methods can be cross-checked. The rest of this paper
 is organized as follows. In Sec.~\ref{sec2}, we briefly
 introduce the Pantheon SNIa sample. In Secs.~\ref{sec3}, \ref{sec4}
 and \ref{sec5}, the cosmic anisotropy is tested with the HC method,
 the DF method, and HEALPix, respectively. In Sec.~\ref{sec6}, some
 brief concluding remarks are given.

\vspace{4mm}{\bf Note added:\ }When most of our computations
 have been completed, we were aware of a similar work by Sun
 and Wang~\cite{Sun:2018cha} (SW18 hereafter) submitted to
 arXiv few days ago. They also tested the cosmic anisotropy
 in the Pantheon SNIa sample by using both the HC method and the DF
 method. Our work is distinct from SW18~\cite{Sun:2018cha} in
 (at least) five aspects: (1) We found the true preferred direction
 with the maximum anisotropy level much higher than the
 one found in SW18 by using the HC method.
 (2)~We have taken the systematic covariance matrix into
 account when using the Pantheon data, while SW18 only
 considered the statistical uncertainty and ignored the
 systematic covariance matrix.
 (3) We checked the results with the simulated isotropic data
 in a way different from the one of SW18.
 (4) We found a result different from the one of SW18, by using
 the DF method, mainly due to the different Markov Chain
 Monte Carlo (MCMC) algorithms used in our work and SW18.
 (5) We further tested the cosmic anisotropy with
 HEALPix, which has not been considered in SW18.
 There exist other minor differences in details. We will
 briefly point out the differences between our work and SW18
 throughout this paper. We stress that our work is completely
 independent, and is complementary to SW18.


 \begin{center}
 \begin{figure}[tb]
 \centering
 \vspace{-3mm}  
 \includegraphics[width=0.93\textwidth]{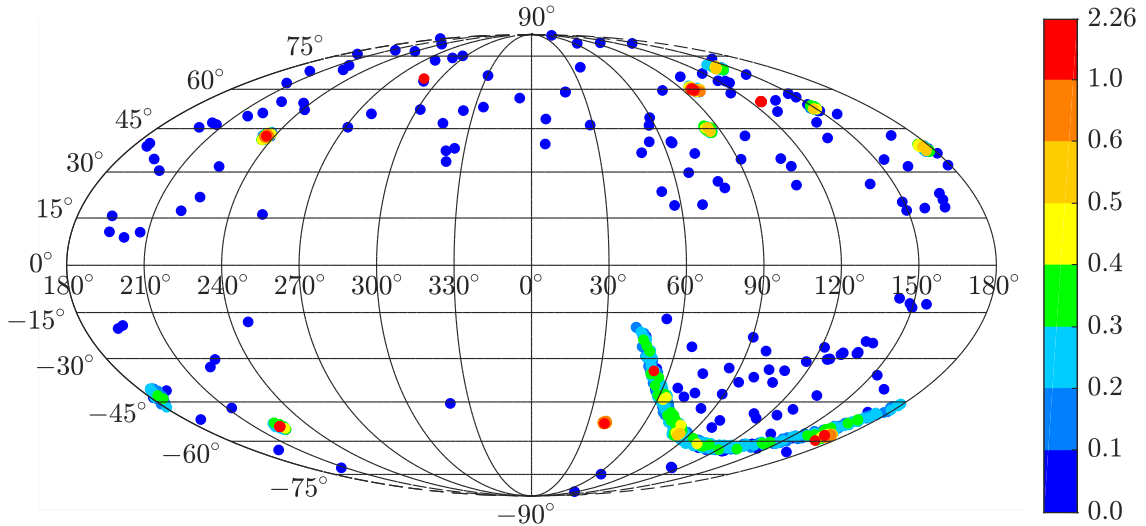}
 \caption{\label{fig1} The distribution of 1048 Pantheon SNIa
 in the galactic coordinate system, while the pseudo-colors
 indicate the redshifts of these SNIa. See the text and
 \cite{disSW18} for details.}
 \end{figure}
 \end{center}


\vspace{-9mm}  


\section{The Pantheon SNIa sample}\label{sec2}

The Pantheon sample~\cite{Scolnic:2017caz,Pantheondata,
 Pantheonplugin} consists of 1048 SNIa. The redshift range of these
 1048 SNIa is given by $0.01012\leq z\leq 2.26$. In Fig.~\ref{fig1},
 we show the distribution of 1048 Pantheon SNIa in the galactic
 coordinate system (see however~\cite{disSW18}), while the
 pseudo-colors indicate the redshifts of these SNIa. It is easy
 to see that most of 1048 Pantheon SNIa are at low redshifts. On the
 other hand, more than one half of 1048 Pantheon SNIa are located in
 the right-bottom region of Fig.~\ref{fig1} (see
 also Sec.~\ref{sec5} for details).

Instead of using the full set of parameters in the Pantheon dataset
 which are extremely complicated, as a convenient alternative,
 one can use the Pantheon plugin~\cite{Pantheonplugin} for
 CosmoMC~\cite{Lewis:2002ah} to constrain cosmological models.
 CosmoMC is configured to expect SNIa data formatted for a
 SALT2 fit~\cite{Conley:2011ku}, as in the case of the JLA
 sample~\cite{Betoule:2014frx}), and to perform the SALT2 fit
 simultaneously with other cosmological constraints. The
 SALT2 model is an empirical model for the band correction,
 \be{reveq1}
 \Delta \mu_B= \left(m-M\right) - \left(m_B - M_B\right)\,,
 \ee
 that relates the actual observed $B$-band apparent magnitudes,
 $m_B$, to the inferred bolometric apparent magnitudes, $m$,
 where $M_B$ and $M$ are the absolute $B$-band magnitudes and
 absolute bolometric magnitudes, respectively. The theoretical
 distance modulus is a bolometric one (over all frequencies),
 which is why the band correction is required. The SALT2
 model~\cite{Conley:2011ku} is linear in the stretch parameter,
 $x_1$, and the color parameter, $c$, of the SNIa light curves
 \be{DeltaSALT}
 \Delta \mu_B= \alpha x_1 - \beta c\,,
 \ee
 where both the coefficients $\alpha$ and $\beta$ are global
 nuisance parameters to be determined. This leads
 to the corrected bolometric distance modulus
 \be{SALT}
 \mu_{\rm obs}\equiv m-M = m_B - M_B + \alpha x_1 - \beta c\,.
 \ee

The Pantheon dataset has been reduced by the BBC
 method~\cite{bbc} in which the SALT2 corrections~(\ref{SALT})
 are supplemented by additional corrections, namely
 \be{Tripp}
 \mu_{\rm obs}\equiv m-M = m_B - M_B +\alpha x_1 -\beta c
 +\Delta_M+\Delta_B\,,
 \ee
 $\Delta_M$ being a distance correction based on the host
 galaxy rest mass and $\Delta_B$ a distance correction from
 various biases predicted from simulations. Although the BBC
 method is not implemented in CosmoMC, one can still use CosmoMC by
 accepting the corrected apparent bolometric magnitudes
 $m=\mu_{\rm obs}+M$ as determined by Scolnic
 {\it et al.}~\cite{Scolnic:2017caz,Pantheondata}. As a
 computational heuristic~\cite{Pantheonplugin}, one sets
 $\alpha=\beta=0$ in CosmoMC (although this is of course not true of
 the actual fitted values~\cite{Scolnic:2017caz}), and one then
 supplies the corrected bolometric apparent magnitudes in place
 of the actual $B$-band magnitudes expected by the CosmoMC. As
 far as tests of anisotropy of the Hubble law are concerned,
 this means that any signature of anisotropy that is degenerate with
 any of the parameters in the BBC fitting will not be probed by the
 analysis we adopt~\cite{revfoot}.

The theoretical distance modulus predicted by the cosmological model
 is defined by~\cite{Weinberg,Betoule:2014frx}
 \be{eq2}
 \mu_{\rm mod}=5\log_{10}\frac{d_L}{\rm Mpc}+25\,,
 \ee
 where $d_L=\left(v_c/H_0\right)D_L$ is the luminosity
 distance, $v_c$ is the speed of light, $H_0$ is the Hubble constant,
 \be{eq3}
 D_L\equiv(1+z_{\rm hel})\int_0^{z_{\rm cmb}}
 \frac{d\tilde{z}}{E(\tilde{z})}\,,
 \ee
 in which $E(z)\equiv H(z)/H_0$ is the dimensionless Hubble
 parameter, and $z_{\rm cmb}$, $z_{\rm hel}$ are the CMB
 frame redshift and heliocentric redshift, respectively.
 We use the Pantheon plugin~\cite{Pantheonplugin}, with the
 corrected bolometric apparent magnitudes $m=\mu_{\rm obs}+M$,
 as determined from Eq.~(\ref{Tripp}) by the BBC method. This
 leads to
 \be{reveq2}
 \chi^2_{\rm Pan}=\Delta{\boldsymbol{\mu}}^{\,T}
 \cdot\boldsymbol{C}^{-1}\cdot\Delta{\boldsymbol{\mu}}=
 \Delta\boldsymbol{m}^{\,T}\cdot
 \boldsymbol{C}^{-1}\cdot\Delta\boldsymbol{m}\,,
 \ee
 where for the $i$-th
 SNIa, $\Delta\mu_i=\mu_{{\rm obs},i}-\mu_{{\rm mod},i}$
 or $\Delta m_i=m_i-m_{{\rm mod},i}\,$, with
 \be{reveq3}
 m_{\rm mod} = 5\log_{10} D_L + {\cal M}\,,
 \ee
 $\cal M$ being a nuisance parameter corresponding to some
 combination of the absolute magnitude, $M$, and the Hubble
 constant, $H_0$, while the total covariance matrix $\boldsymbol{C}$
 is given by~\cite{Scolnic:2017caz}
 \be{reveq4}
 \boldsymbol{C}=\boldsymbol{D}_{\rm stat}+\boldsymbol{C}_{\rm sys}\,.
 \ee
 Here $\boldsymbol{D}_{\rm stat}$ is the diagonal covariance matrix
 of the statistical uncertainties obtained by adding in quadrature the
 uncertainties associated with the BBC method according to
 Eq.~(4) of~\cite{Scolnic:2017caz}, while $\boldsymbol{C}_{\rm sys}$
 is the covariance matrix of systematic uncertainties in
 the BBC approach, which differs somewhat from the SALT2
 approach~\cite{Conley:2011ku} since the BBC method produces
 distances from the fit parameters directly. The entries
 \be{reveq5}
 \boldsymbol{D}_{{\rm stat},ii}=\sigma^2_{m_i}\,,
 \ee
 along with the corrected bolometric apparent magnitudes, $m_i$, and
 the CMB rest-frame and heliocentric rest-frame redshifts,
 $z_{{\rm cmb},i}$ and $z_{{\rm hel},i}\,$, are given by the
 data file {\tt lcparam$_{-}$full$_{-}$long.txt} in the
 Pantheon plugin~\cite{Pantheonplugin} for CosmoMC, while
 $\boldsymbol{C}_{\rm sys}$ is given in the file
 {\tt sys$_{-}$full$_{-}$long.txt}~\cite{Pantheonplugin}. Since
 $H_0$ is absorbed into $\cal M$ in the analytic marginalization, the
 Pantheon sample is independent of the normalization of the
 Hubble constant.
 According to the setting file {\tt full$_{-}$long.dataset} in
 the Pantheon plugin~\cite{Pantheonplugin} for CosmoMC, the
 nuisance parameter $\cal M$ is marginalized by using Eq.~(C1)
 in Appendix~C of~\cite{Conley:2011ku} (note that in the JLA
 dataset $\cal M$ is marginalized by using
 Eq.~(C2) of~\cite{Conley:2011ku}
 instead~\cite{Wang:2015tua,Zou:2017ksd}). The best-fit model
 parameters (and their uncertainties) can be obtained by
 minimizing $\chi^2_{\rm Pan}$.

\vspace{4mm}{\bf Note added:\ }In SW18~\cite{Sun:2018cha}, they
 only considered the statistical uncertainties, and actually
 ignored the systematic covariance matrix $\boldsymbol{C}_{\rm sys}$.
 In the present work, we instead use the full covariance matrix
 with the systematic uncertainties determined by the BBC
 method~\cite{bbc}, rather than
 simply the statistical uncertainties as in SW18.


 \begin{center}
 \begin{figure}[tb]
 \centering
 \vspace{-6mm}  
 \includegraphics[width=0.93\textwidth]{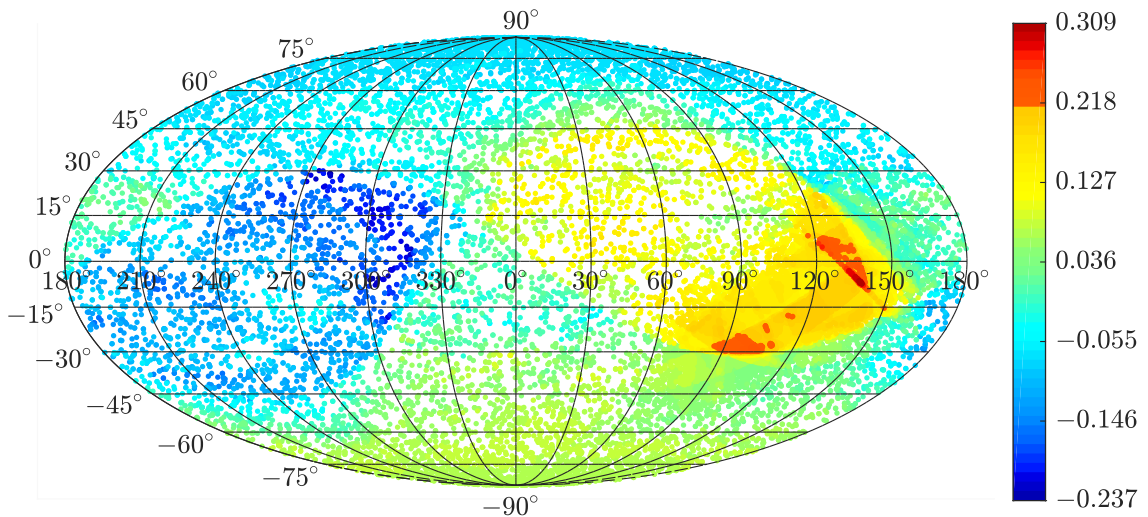}
 \caption{\label{fig2} The pseudo-color map of AL$(l,\,b)$ in
 the galactic coordinate system, obtained by using the HC method to
 the Pantheon SNIa sample. The two ``preferred directions''
 $(138.08^\circ,\,-6.84^\circ)$ and $(102.36^\circ,\,-28.58^\circ)$
 are within the red regions. See the text for details.}
 \end{figure}
 \end{center}


\vspace{-10mm}  


\section{Testing the cosmic anisotropy with the HC method}\label{sec3}

In the present work, we test the cosmic anisotropy in the
 Pantheon SNIa sample by using three methods, and hence the
 results from different methods can be cross-checked. For
 simplicity, we consider the spatially flat $\Lambda$CDM
 model throughout this work. As is well known, in this model,
 the dimensionless Hubble parameter is given by
 \be{eq10}
 E(z)=\left[\,\Omega_{m0}(1+z)^3+
 \left(1-\Omega_{m0}\right)\,\right]^{1/2}\,,
 \ee
 where $\Omega_{m0}$
 is the fractional density of the pressureless matter.

At first, we consider the HC method proposed
 in~\cite{Schwarz:2007wf} and then improved
 by~\cite{Antoniou:2010gw}. As is well known, the deceleration
 parameter $q_0=-1+3\Omega_{m0}/2$ in the flat $\Lambda$CDM
 model. So, it is convenient to use $\Omega_{m0}$ instead of
 $q_0$ to characterize the cosmic acceleration~\cite{Antoniou:2010gw}.
 Following~\cite{Antoniou:2010gw} (and e.g.~\cite{Cai:2011xs,
 Yang:2013gea,Chang:2014nca,Deng:2018yhb}), the main steps to
 implement the HC method are ({\it i}) Generate a random
 direction $\hat{r}_{\rm rnd}$ indicated by $(l,\,b)$ with a uniform
 probability distribution, where $l\in [\,0^\circ,\,360^\circ)$ and
 $b\in [-90^\circ,\,+90^\circ\,]$ are the longitude and the
 latitude in the galactic coordinate system, respectively.
 ({\it ii}) Divide the Pantheon dataset into two subsets
 according to the sign of the inner product
 $\hat{r}_{\rm rnd}\cdot\hat{r}_{\rm dat}$, where
 $\hat{r}_{\rm dat}$ is a unit vector describing the direction
 of each SNIa in the Pantheon dataset. So, one subset corresponds to
 the hemisphere in the direction of the random vector (defined
 as ``up''), while the other subset corresponds to the opposite
 hemisphere (defined as ``down''). Noting that the position of
 each SNIa in the Pantheon sample~\cite{Panradec} is given
 by right ascension (ra) and declination (dec) in degree
 (equatorial coordinate system, J2000), one should convert
 $\hat{r}_{\rm rnd}$ and $\hat{r}_{\rm dat}$ to Cartesian
 coordinates in this step. ({\it iii}) Find the best-fit values
 on $\Omega_{m0}$ in each hemisphere ($\Omega_{m0,u}$ and
 $\Omega_{m0,d}$), and then obtain the so-called anisotropy
 level (AL) quantified through the normalized
 difference~\cite{Antoniou:2010gw}, namely
  \be{eq11}
 {\rm AL}\equiv\frac{\Delta\Omega_{m0}}{\bar{\Omega}_{m0}}=
 2\cdot\frac{\Omega_{m0,u}-\Omega_{m0,d}}{\Omega_{m0,u}
 +\Omega_{m0,d}}\,.
 \ee
 ({\it iv}) Repeat for $N$ random directions
 $\hat{r}_{\rm rnd}$ and find the maximum AL, as well as the
 corresponding direction of maximum anisotropy.
 ({\it v}) Obtain the $1\sigma$ uncertainty $\sigma_{\rm AL}$
 associated with the maximum AL~\cite{Antoniou:2010gw},
 \be{eq12}
 \sigma_{\rm AL}=\frac{\sqrt{\sigma_{\Omega_{m0,u}^{\rm max}}^2
 +\sigma_{\Omega_{m0,d}^{\rm max}}^2}}{\Omega_{m0,u}^{\rm max}
 +\Omega_{m0,d}^{\rm max}}\,.
 \ee
 Note in~\cite{Antoniou:2010gw} that $\sigma_{\rm AL}$ is
 the error due to the uncertainties of the SNIa distance
 moduli propagated to the best-fit $\Omega_{m0}$ on each
 hemisphere and thus to AL. One can identify all the test
 axes corresponding to
 ${\rm AL=AL}_{\rm max}\pm\sigma_{\rm AL}$. These axes
 cover an angular region corresponding to the $1\sigma$
 range of the maximum anisotropy direction. We
 refer to~\cite{Antoniou:2010gw} for more details
 of the HC method.

Let us implement the HC method to the Pantheon
 SNIa sample~\cite{Scolnic:2017caz,Pantheondata,Pantheonplugin,
 Panradec}. First, we repeat 15000 random directions $(l,\,b)$
 across the whole sky, and find that the directions with the
 largest ALs concentrate around two directions
 $(137^\circ,\,-7^\circ)$ and $(110^\circ,\,-20^\circ)$. Then,
 we densely repeat 30000 random directions around these two
 preliminary directions. Finally, we find that the $1\sigma$
 angular region with the maximum AL is in the direction
 \be{eq13}
 (l,\,b)\,_{\rm HC,\,max}^{\rm Pan}
 =({138.0841^\circ}\,^{+3.1574^\circ}_{-16.9046^\circ}\,,\,
 {-6.8407^\circ}\,^{+13.5501^\circ}_{-2.3119^\circ})\,,
 \ee
 and the corresponding maximum AL (with $1\sigma$
 uncertainty) is
 \be{eq14}
 {\rm AL_{max}^{Pan}}=0.3088\pm 0.0738\,.
 \ee
 In addition, we also find a sub-maximum AL in the direction
 (with $1\sigma$ uncertainty)
 \be{eq15}
 (l,\,b)\,_{\rm HC,\,sub}^{\rm Pan}
 =({102.3584^\circ}\,^{+47.9475^\circ}_{-34.2187^\circ}\,,\,
 {-28.5775^\circ}\,^{+50.6001^\circ}_{-1.7822^\circ})\,,
 \ee
 and the corresponding sub-maximum AL (with $1\sigma$ uncertainty) is
 \be{eq16}
 {\rm AL_{sub}^{Pan}}=0.2411\pm 0.0710\,.
 \ee
 In fact, it is not so rare to find two preferred directions
 (see e.g.~\cite{Zhou:2017lwy,Deng:2018yhb}). We present the
 pseudo-color map of AL$(l,\,b)$ in Fig.~\ref{fig2}. It is
 clear to see these two preferred directions within the red
 regions. However, it is worth noting that these two directions are
 not in agreement with most of the preferred directions
 found in various observational datasets (see Table~I and
 Fig.~10 of~\cite{Deng:2018yhb}). This unusual fact makes these
 results impeachable.

So, it is of interest to see whether the maximum ALs in
 Eqs.~(\ref{eq14}) and (\ref{eq16}) for the real Pantheon data
 are consistent with statistical isotropy. To this end, we can
 compare the real data with the simulated isotropic data
 following~\cite{Antoniou:2010gw}. The idea to construct the
 simulated isotropic data is simple. First, we fit the flat
 $\Lambda$CDM model to the real Pantheon data, and get the
 best-fit model parameter $\Omega_{m0}=0.298111$, while the
 corresponding ${\cal M}=23.808891$
 (see Appendix~C of~\cite{Conley:2011ku} for technical
 details). Then, except that the value of $m_i$ should be
 simulated, we keep all the other numerical observed data
 unchanged for each SNIa in the Pantheon sample. We obtain
 $m_{{\rm mod},i}$ by using Eqs.~(\ref{reveq3})
 and (\ref{eq3}), (\ref{eq10}) with the best-fit $\Omega_{m0}$
 and $\cal M$ mentioned above for each SNIa. Finally, the
 simulated $m_i$ can be obtained by taking a random number from
 a Gaussian distribution with the mean at the corresponding
 $m_{{\rm mod},i}$, while the standard deviation of this
 Gaussian distribution is equal to the observed $\sigma_{m_i}$.
 So far, a simulated isotropic dataset consisting of 1048 SNIa
 is ready. We can then compare the absolute maximum AL of the simulated
 isotropic dataset with the one of the real Pantheon dataset, which
 are both found by repeating 1000 random directions in the
 corresponding dataset. We generate 100 simulated isotropic datasets
 and make such kind of comparisons for 100 times. In fact,
 $100\times 1000$ computations consume a lot of computing power and
 time, while they have acceptable statistics. Finally, we find that
 the absolute maximum AL of the real Pantheon dataset is larger
 (smaller) than the one of the simulated isotropic dataset for
 55~(45) times, respectively. This is not statistically significant
 in fact. Thus, the maximum AL of the Pantheon dataset is
 consistent with statistical isotropy. No evidence for the
 cosmic anisotropy is found.

\vspace{4mm}{\bf Note added:\ }In SW18~\cite{Sun:2018cha}, they
 used only 500 random directions to search the maximum AL in
 the Pantheon sample, and hence the ``maximum'' AL they found
 is only $0.105$, which is much smaller than $0.3088\,$ found
 in the present work by using 45000 random directions at the
 price of great computing power and time. On the other hand, we
 checked the results with the simulated isotropic data in a way
 slightly different from the one of SW18~\cite{Sun:2018cha}.


 \begin{center}
 \begin{figure}[tb]
 \centering
 \vspace{-5mm}  
 \includegraphics[width=0.9\textwidth]{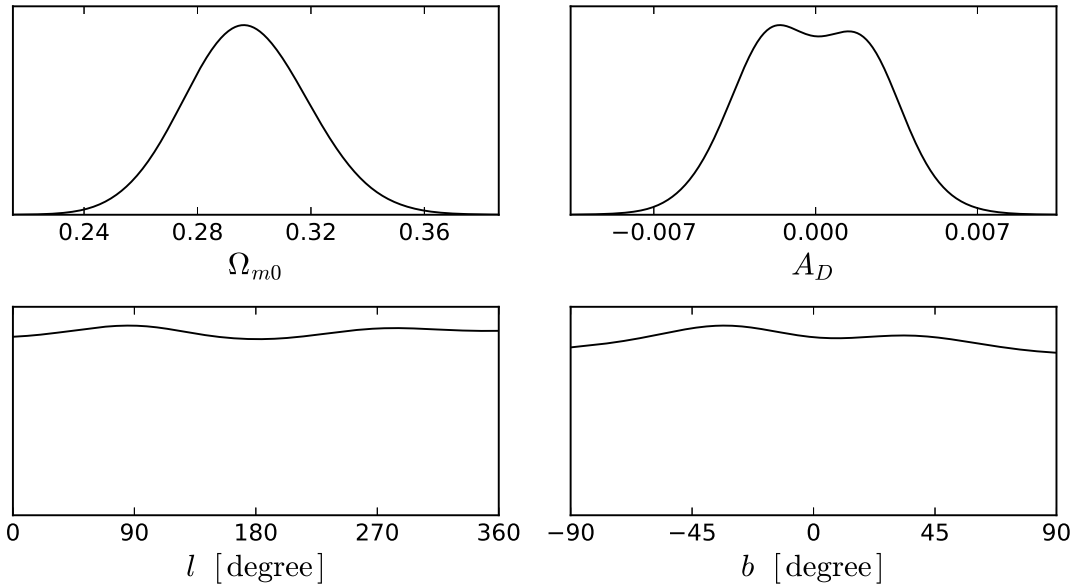}
 \caption{\label{fig3} The marginalized probability distributions of
 $\Omega_{m0}$, the dipole magnitude $A_D$, and the dipole direction
 $(l,\,b)$, obtained by using the DF method to the Pantheon
 data. See the text for details.}
 \end{figure}
 \end{center}


\vspace{-10mm}  


\section{Testing the cosmic anisotropy with the DF method}\label{sec4}

Let us use another method to test the cosmic anisotropy in the
 Pantheon data, and cross-check the result with the one obtained in
 the previous section. The DF method was proposed
 in~\cite{Mariano:2012wx}, and it has been extensively considered in
 the literature (e.g.~\cite{Mariano:2012wx,Chang:2014nca,Lin:2015rza,
 Chang:2017bbi,Zhou:2017lwy,Chang:2018vxs,Yang:2013gea,Wang:2014vqa,
 Deng:2018yhb}). As mentioned in Sec.~\ref{sec2}, the observational
 quantity under consideration is the corrected bolometric
 magnitude $m$ in the Pantheon plugin~\cite{Pantheonplugin}.
 If it is anisotropic, we can consider a dipole correction,
 and replace $m_{\rm mod}$ with
 \be{eq17}
 m_{\rm mod}=\bar{m}_{\rm mod}\left[\,1+
 A_D\left(\hat{n}\cdot\hat{p}\right)\,\right].
 \ee
 where $\bar{m}_{\rm mod}$ is the value predicted by the
 isotropic theoretical model given by Eq.~(\ref{reveq3}), and $A_D$ is
 the dipole magnitude, $\hat{n}$ is the dipole direction, $\hat{p}$
 is the unit 3-vector pointing towards the data point. In terms
 of the galactic coordinates $(l,\,b)$, the dipole direction is
 given by
 \be{eq18}
 \hat{n}=\cos(b)\cos(l)\,\hat{\bf i}+\cos(b)
 \sin(l)\,\hat{\bf j}+\sin(b)\,\hat{\bf k}\,,
 \ee
 where $\hat{\bf i}$, $\hat{\bf j}$, $\hat{\bf k}$ are the
 unit vectors along the axes of Cartesian coordinates system.
 The position of the $i$-th data point with the
 galactic coordinates $(l_i,\,b_i)$ is given by
 \be{eq19}
 \hat{p}_i=\cos(b_i)\cos(l_i)\,\hat{\bf i}+
 \cos(b_i)\sin(l_i)\,\hat{\bf j}+\sin(b_i)\,\hat{\bf k}\,.
 \ee
 One can find the best-fit dipole direction $(l,\,b)$ and the
 dipole magnitude $A_D$ as well as the other model parameters
 by minimizing the corresponding $\chi^2$. In doing this, the
 Markov Chain Monte Carlo (MCMC) code CosmoMC~\cite{Lewis:2002ah} is
 used, and the nuisance parameters $\cal M$ can
 be marginalized~\cite{Betoule:2014frx}. Note that the Pantheon
 plugin for CosmoMC is available at~\cite{Pantheonplugin}. Here, we
 let $A_D$ be a completely free parameter. In Fig.~\ref{fig3},
 we show the marginalized probability distributions of
 $\Omega_{m0}$, the dipole magnitude $A_D$ and the dipole direction
 $(l,\,b)$. The constraints with $1\sigma$ uncertainties are given by
 \bea
 &\Omega_{m0}=0.2975^{+0.0219}_{-0.0218}\,,~~~~~~~
 A_D=(-0.0931^{+2.9619}_{-2.9316})\times 10^{-3}\,,\label{eq20}\\[2mm]
 &0^\circ\leq l\leq 360^\circ\,,~~~~~~~
 -90^\circ\leq b\leq 90^\circ\,. \label{eq21}
 \eea
 It is clear to see that $A_D=0$ is fully consistent with the
 Pantheon data, and the $1\sigma$ region of $l$ and $b$ is the
 whole sky. No evidence for the cosmic anisotropy is found.

\begin{center}
 \begin{figure}[tb]
 \centering
 \vspace{-6mm}  
 \includegraphics[width=0.9\textwidth]{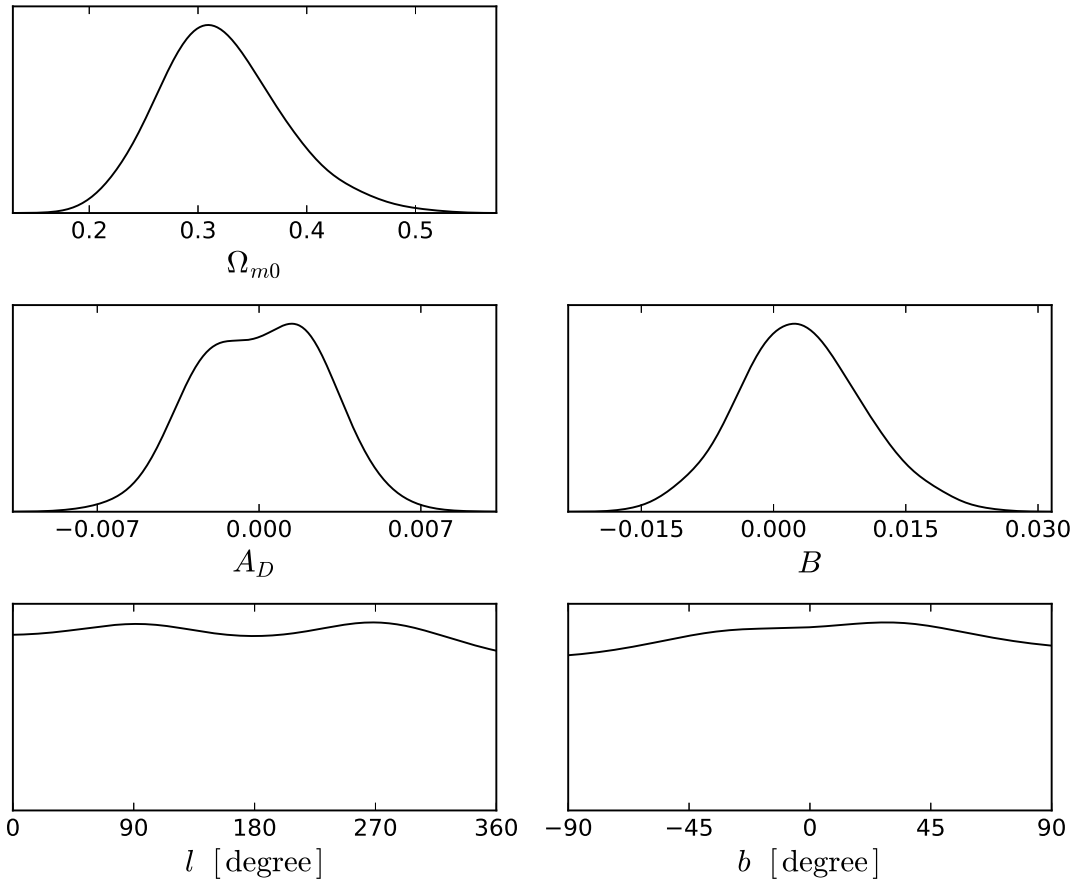}
 \caption{\label{fig4} The marginalized probability distributions of
 $\Omega_{m0}$, the dipole magnitude $A_D$, the monopole term
 $B$, and the dipole direction $(l,\,b)$, obtained by using the
 DF method to the Pantheon data. See the text for details.}
 \end{figure}
 \end{center}

One can generalize Eq.~(\ref{eq17}) by including the monopole
 term $B$ in addition, namely
 \be{eq22}
 m_{\rm mod}=\bar{m}_{\rm mod}\left[\,1+B
 +A_D\left(\hat{n}\cdot\hat{p}\right)\,\right].
 \ee
 Similarly, in Fig.~\ref{fig4}, we show the marginalized probability
 distributions of $\Omega_{m0}$, the dipole magnitude $A_D$,
 the monopole term $B$, and the dipole direction $(l,\,b)$.
 The constraints with $1\sigma$ uncertainties are given by
 \bea
 &\Omega_{m0}=0.3226^{+0.0486}_{-0.0649}\,,~~~~~
 A_D=(0.0680^{+2.8751}_{-2.9435})\times 10^{-3}\,,~~~~~
 B=(3.1628^{+6.6875}_{-7.6078})\times 10^{-3}\,,\label{eq23}\\[2mm]
 &0^\circ\leq l\leq 360^\circ\,,~~~~~~~
 -90^\circ\leq b\leq 90^\circ\,. \label{eq24}
 \eea
 Again, it is clear to see that $A_D=0$ and $B=0$ are fully
 consistent with the Pantheon data, and the $1\sigma$ region of
 $l$ and $b$ is the whole sky. No evidence for the cosmic
 anisotropy is found.

So, by using the DF method, we confirm the result obtained by
 using the HC method in the previous section, namely there is
 no cosmic anisotropy in the Pantheon data.

\vspace{4mm}{\bf Note added:\ }In SW18~\cite{Sun:2018cha}, they
 used a fairly different MCMC algorithm by their own, and
 obtained fairly different results. In particular, they found a
 relatively small $1\sigma$ region of the dipole direction, as
 was shown in Fig.~4 of SW18~\cite{Sun:2018cha}. Instead, we
 used the MCMC code CosmoMC~\cite{Lewis:2002ah,Pantheonplugin},
 and found that the $1\sigma$ region of the dipole direction is
 the whole sky.


 \begin{center}
 \begin{figure}[tb]
 \centering
 \vspace{-10mm}  
 \includegraphics[width=0.55\textwidth]{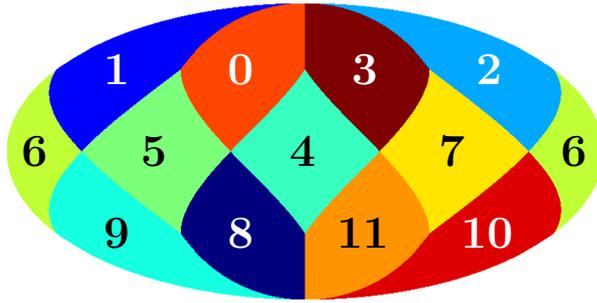}
 \caption{\label{fig5} The whole sky is divided into 12
 equal-area regions by using HEALPix, which are labeled from
 0 to 11. Note that here we use the same galactic coordinate
 system as in Fig.~\ref{fig1}, rather than the default one of
 HEALPix. See the text for details.}
 \end{figure}
 \end{center}



 \begin{table}[tb]
 \renewcommand{\arraystretch}{1.5}
 \begin{center}
 \vspace{4mm} 
 \begin{tabular}{lllllll} \hline\hline

 Region\hspace{7mm}  &    $0$ &  \hspace{-12mm}  $1$ &  $2$ &  $3$ &    $4$ &  \hspace{5mm} $5$  \\ \hline
 $N_{\rm SN}$   &  $16$ &  \hspace{-12mm}  $112$ &    $191$ &  $49$ &    $0$ &  \hspace{5mm} $2$   \\
 $\Omega_{m0}$   &   N/A    &  \hspace{-12mm}  $0.2633^{+0.0362}_{-0.0411} $ &    $0.2922^{+0.0311}_{-0.0349} $ &    $0.3174^{+0.0771}_{-0.0903} $ &   N/A    & \hspace{5mm} N/A \\
 $q_0$   &  N/A  &  \hspace{-12mm} $-0.6051^{+0.0543}_{-0.0615} $ \hspace{5mm} &   $-0.5618^{+0.0467}_{-0.0522} $ \hspace{5mm} &  $-0.5239^{+0.1157}_{-0.1355} $ \hspace{7mm} &  N/A  & \hspace{5mm} N/A \\

\hline\hline

 Region  &    $6$ &    $7$ &  \hspace{-5mm} $8$ &  \hspace{-19mm} $9$ &  \hspace{-19mm} $10$ &    $11$ \  \\  \hline
 $N_{\rm SN}$   &    $34$ &    $7$ &  \hspace{-5mm} $1$ &  \hspace{-19mm} $69$ &  \hspace{-19mm} $303$ &    $264$ \  \\
 $\Omega_{m0}$   &   $0.1566^{+0.0667}_{-0.1062} $ & N/A  &  \hspace{-5mm} N/A & \hspace{-19mm} $0.3585^{+0.0542}_{-0.0663} $ &  \hspace{-19mm} $0.2788^{+0.0297}_{-0.033} $ &    $0.3139^{+0.0331}_{-0.0369} $ \\
 $q_0$   &  $-0.7651^{+0.0992}_{-0.1594} $ \hspace{5mm} & N/A & \hspace{-5mm} N/A & \hspace{-19mm} $-0.4623^{+0.0814}_{-0.0994} $ &  \hspace{-19mm} $-0.5819^{+0.0446}_{-0.0494} $ \hspace{5mm} &  $-0.5292^{+0.0496}_{-0.0554}$ \\

 \hline\hline
 \end{tabular}
 \end{center}
 \caption{\label{tab1} The number of SNIa, and the best-fit
 $\Omega_{m0}$, $q_0$ with $1\sigma$ uncertainties for each region.}
 \end{table}


\vspace{-19mm}  


\section{Testing the cosmic anisotropy with HEALPix}\label{sec5}

Last, we consider the third method to test the cosmic anisotropy in
 the Pantheon data. In the HC method, the directions are compared in
 the way of ``up hemisphere'' versus ``down hemisphere''. So,
 the fine structure might be smoothed in fact. Actually, in
 Sec.~2 of~\cite{Antoniou:2010gw} (one of the original papers
 proposed the HC method), the authors mentioned this issue and
 referred to HEALPix~\cite{Gorski:2004by} as a solution.
 In e.g.~\cite{Zhao:2013yaa}, the cosmic anisotropy in the
 Union2 SNIa dataset was tested with HEALPix.
 Following~\cite{Zhao:2013yaa}, here we use
 it to the Pantheon SNIa dataset.

HEALPix~\cite{Gorski:2004by} is a genuinely curvilinear
 partition of the sphere into exactly equal area quadrilaterals
 of varying shape. The base-resolution comprises 12 pixels in
 three rings around the poles and equator. The resolution of
 the grid is expressed by the parameter $N_{\rm side}$ which
 defines the number of divisions along the side
 of a base-resolution pixel that is needed to reach a desired
 high-resolution partition~\cite{Gorski:2004by}.
 Following~\cite{Zhao:2013yaa}, we adopt $N_{\rm side}=1$,
 namely the lowest resolution (12 pixels), due to the fact that
 the number of SNIa in the dataset under consideration is not
 large enough. In this case, the whole sky is divided into 12
 equal-area regions by using HEALPix, which are labeled from
 0 to 11, as is shown in Fig.~\ref{fig5}. Note that here we use
 the same galactic coordinate system as in Fig.~\ref{fig1},
 rather than the default one of HEALPix. This can be done by
 using a suitable coordinate transformation.


 \begin{center}
 \begin{figure}[tb]
 \centering
 \vspace{-2mm}  
 \includegraphics[width=1.0\textwidth]{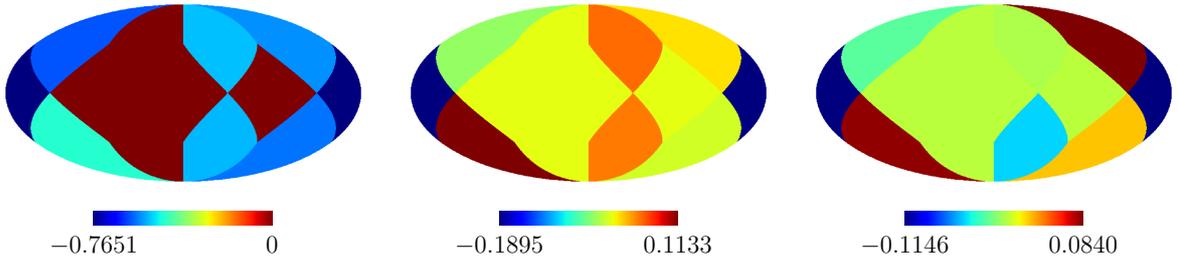}
 \caption{\label{fig6} Left panel: The pseudo-color map of the
 best-fit $q_0$ in the regions 1, 2, 3, 6, 9, 10, 11, while the
 regions 0, 4, 5, 7, 8 are masked. Middle panel: The monopole
 is subtracted from Left panel. Right panel: Both the monopole
 and the dipole are subtracted from Left panel. See the
 text for details.}
 \end{figure}
 \end{center}


\vspace{-10.8mm}  

The distribution of 1048 Pantheon SNIa in the galactic
 coordinate system is shown in Fig.~\ref{fig1}. Using the
 routine {\tt ang2pix\,*} provided in HEALPix package,
 we can find the corresponding SNIa in each region. The numbers
 of SNIa in all the 12 regions are given in Table~\ref{tab1}.
 Obviously, the numbers of SNIa in the regions 0, 4, 5, 7, 8
 are $N_{\rm SN}=16$, 0, 2, 7, 1, respectively. There are too
 few SNIa in these five regions. Therefore, we certainly
 exclude them from the following discussions. Note that the
 region 6 is subtle. There are 34 SNIa in the region 6. Not too
 many, not too few. Thus, we will consider both cases with and
 without the region 6. We can constrain the model parameter
 $\Omega_{m0}$ and the derived deceleration
 parameter $q_0=-1+3\Omega_{m0}/2$ by fitting the flat
 $\Lambda$CDM model to the corresponding SNIa in each region.
 In Table~\ref{tab1}, we also present the best-fit
 $\Omega_{m0}$, $q_0$ with $1\sigma$ uncertainties for each
 region. Following~\cite{Zhao:2013yaa}, we use $q_0$ as the
 diagnostic for the anisotropy of the cosmic acceleration.

At first, we take the region 6 into consideration. In this
 case, we mask the five regions 0, 4, 5, 7, 8 mentioned
 above (and flag them as bad pixels). In the left panel of
 Fig.~\ref{fig6}, we show the best-fit $q_0$ in each region,
 and flag the masked regions by $q_0\sim 0$. It is convenient
 to expand the 2D anisotropic $q_0$ map in spherical harmonics,
 namely to consider the multipole
 expansion~\cite{Gorski:2004by,Multipole},
 \be{eq25}
 q_0\left(\theta,\,\phi\right)=\sum\limits_{l=0}^\infty
 \sum\limits_{m=-l}^l a_l^m Y_l^m\left(\theta,\,\phi\right)\,,
 \ee
 where $Y_l^m\left(\theta,\,\phi\right)$ are the standard spherical
 harmonics, and $a_l^m$ are constant coefficients which depend
 on the function. The term $a_0^0$ represents the monopole, the
 terms $a_1^{-1}$, $a_1^0$, $a_1^1$ represent the dipole, and so on.
 Equivalently, this series is also frequently
 written as~\cite{Multipole,Thompson:1994}
 \be{eq26}
 q_0\left(\theta,\,\phi\right)=a_0+a_i n^i+a_{ij}n^i n^j+
 a_{ijk} n^i n^j n^k+\dots\,,
 \ee
 where $n^i$ represent the components of a unit vector in the
 direction given by the angles $\theta$ and $\phi$, and indices
 are implicitly summed. The term $a_0$ is the monopole, $a_i$
 is a set of three numbers representing the dipole, and so on.
 In this work, we only consider the lowest multipoles, namely
 the monopole with $l=0$ and the dipole with $l=1$, due to the
 low resolution of the anisotropic map. Using the routine
 {\tt remove$_{-}$dipole\,*} provided in HEALPix package, one
 can fit and remove the monopole and the dipole from a HEALPix
 map. Note that the masked (and bad) pixels will not be used
 for fit. First, we try to remove only the monopole from the
 left panel of Fig.~\ref{fig6}. The best-fit monopole is $-0.5756$,
 which is equivalent to the average deceleration parameter in
 the whole sky. If we subtract this best-fit monopole from the
 $q_0$ map, as is shown in the middle panel of Fig.~\ref{fig6}, the
 residual becomes modest (and much smaller than the monopole). Then,
 we try to remove both the monopole and the dipole from the
 left panel of Fig.~\ref{fig6}. The corresponding best-fit monopole
 is $-0.5316$, and the three components of the dipole in the
 HEALPix default Cartesian coordinate system are $0.1189$,
 $6.0550\times 10^{-2}$, $-2.9350\times 10^{-2}$. This dipole
 is much smaller than the monopole in fact. If we subtract
 both the best-fit monopole and dipole from the $q_0$ map, as
 is shown in the right panel of Fig.~\ref{fig6}, the residual
 becomes smaller but keeps the same magnitude as in the middle
 panel. The change between the right and middle panels of
 Fig.~\ref{fig6} is not so significant. The main component in
 the $q_0$ map is contributed by the monopole, and the
 anisotropic component contributed by the dipole is relatively
 small.

\begin{center}
 \begin{figure}[tb]
 \centering
 \vspace{-2mm}  
 \includegraphics[width=1.0\textwidth]{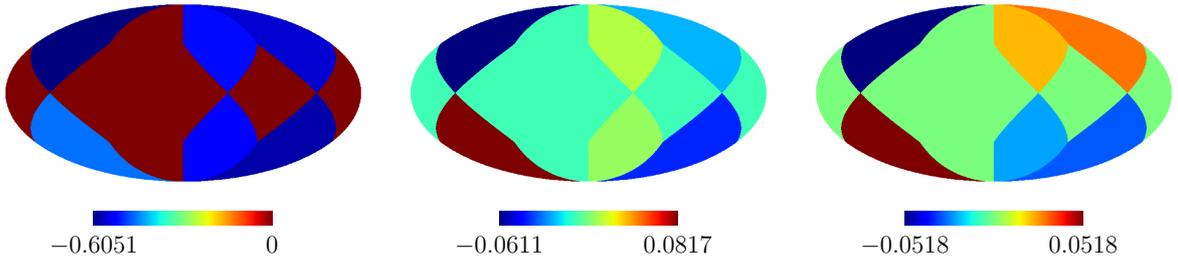}
 \caption{\label{fig7} Left panel: The pseudo-color map of the
 best-fit $q_0$ in the regions 1, 2, 3, 9, 10, 11, while the
 regions 0, 4, 5, 6, 7, 8 are masked. Middle panel: The
 monopole is subtracted from Left panel. Right panel: Both the
 monopole and the dipole are subtracted from Left panel. See
 the text for details.}
 \end{figure}
 \end{center}

As mentioned above, the region 6 is subtle, and it contains
 only 34 SNIa. Due to the paucity of SNIa in this region,
 the uncertainty of $q_0$ is fairly large, as is shown in
 Table~\ref{tab1}. Since we have taken the region~6 into
 consideration above, let us turn to the case excluding
 the region 6. In this case, we mask six regions 0, 4, 5,
 6, 7, 8 (and flag them as bad pixels). In the left panel of
 Fig.~\ref{fig7}, we show the best-fit $q_0$ in each region,
 and flag the masked regions by $q_0\sim 0$. First, we try to
 remove only the monopole from the left panel of
 Fig.~\ref{fig7}. The best-fit monopole is $-0.5440$. If we
 subtract this best-fit monopole from the $q_0$ map, as is
 shown in the middle panel of Fig.~\ref{fig7}, the residual
 becomes very small. This suggests that the $q_0$ map is highly
 isotropic in fact. Then, we try to remove both the monopole
 and the dipole from the left panel of Fig.~\ref{fig7}. The
 corresponding best-fit monopole is $-0.5301$, and the three
 components of the dipole in the HEALPix default Cartesian
 coordinate system are $4.2964\times 10^{-2}$,
 $3.6163\times 10^{-2}$, $-2.9350\times 10^{-2}$.
 The subtracted map is shown in the right panel of Fig.~\ref{fig7},
 which is very close to the middle panel of Fig.~\ref{fig7}.
 In fact, the anisotropic component contributed by the dipole
 is very small.

Briefly, we find that there is no evidence for the cosmic
 anisotropy in the $q_0$ map. This result obtained by using
 HEALPix is fully in agreement with the ones obtained by using
 both the HC method and the DF method in the previous sections.

\vspace{4mm}{\bf Note added:\ }In SW18~\cite{Sun:2018cha}, they
 have not tested the cosmic anisotropy by using HEALPix. On the
 other hand, after the present work has been submitted to arXiv
 and journal, another work~\cite{Andrade:2018eta} appeared, in
 which the possible cosmic anisotropy in the Pantheon sample
 was also studied by using HEALPix. In~\cite{Andrade:2018eta},
 their results confirm that there is null evidence against
 the cosmological principle.


\vspace{-10mm}  


\section{Concluding remarks}\label{sec6}

The cosmological principle assumes that the universe is homogeneous
 and isotropic on cosmic scales. There exist many works testing the
 cosmic homogeneity and/or the cosmic isotropy of the universe
 in the literature. In fact, some observational hints of the cosmic
 anisotropy have been claimed. However, we note that the paucity of
 the data considered in the literature might be responsible for the
 ``found'' cosmic anisotropy. So, it might disappear in a large
 enough sample. Very recently, the Pantheon sample consisting of 1048
 SNIa has been released, which is the largest spectroscopically
 confirmed SNIa sample to date. In the present work, we test
 the cosmic anisotropy in the Pantheon SNIa sample by using
 three methods, and hence the results from different methods can be
 cross-checked. All the results obtained by using the HC method, the
 DF method and HEALPix suggest that no evidence for the cosmic
 anisotropy is found in the Pantheon SNIa sample.

Some remarks are in order. In this work, we only consider the
 spatially flat $\Lambda$CDM model. In fact, one can generalize
 our discussions to other cosmological models, such as $w$CDM,
 CPL models, or model-independent parameterizations like
 cosmography. It is reasonable to expect that our results
 do not change significantly in these generalized cases.

Both the DF method and HEALPix involve the dipole and the monopole.
 However, in the case of the DF method, the quantity under
 consideration is $m$ (which is equivalent to the distance
 modulus or the luminosity distance). In the case of HEALPix,
 the quantity under consideration is the deceleration parameter
 $q_0$ instead.

In almost all of the relevant works, only the dipole and the
 monopole have been considered. In fact, one can further take
 the quadrupole into account. Although it is not easy to deal
 with the quadrupole, this issue deserves future consideration.

In the case of HEALPix, we divide the whole sky into only 12
 equal-area regions. This corresponds to $N_{\rm side}=1$,
 namely the lowest resolution. In fact, the total number of
 pixels is equal to $N_{\rm pix}=12N_{\rm side}^2$. Although
 the number of SNIa in the dataset under consideration might
 be not large enough, it is still interesting to consider
 a higher resolution (e.g. $N_{\rm side}=2$) in the relevant works.

One should be aware of the caveat that our conclusions might
 change if the anisotropic Hubble law in Eq.~(\ref{eq17}) was
 considered at the same time as the standardization of SNIa
 by the BBC method~\cite{bbc}. As is mentioned in
 Sec.~\ref{sec2}, the Pantheon sample actually use the
 complicated BBC method~\cite{bbc} (rather than the SALT2
 method) in the SNIa band correction, as discussed in Sec.~3.5
 of~\cite{Scolnic:2017caz}. If one were to apply the SALT2
 method in the manner assumed in the JLA sample with non-zero
 $\alpha$ and $\beta$, then one could apply the anisotropic
 Hubble law at the same time as fitting for these parameters.
 However, the BBC method is simply so complicated that it is
 not an easy task.

Nonetheless, from Fig.~\ref{fig1}, it is easy to see that the
 low-redshift SNIa at $z<0.2$ are relatively isotropically
 distributed, but the SNIa at $0.2<z<0.4$ are concentrated in
 the southern latitudes with approximately
 $40^\circ <l< 190^\circ$. Since the BBC method is designed to
 deal with some redshift dependent biases among other effects,
 empirically this could lead to degeneracies with the effects
 that we are searching for if the redshift ranges of the sample
 show an anisotropy. So, we consider a redshift tomography of
 the Pantheon data to explore the possible redshift anisotropy.
 We use the DF method in Sec.~\ref{sec4} to find the possible
 dipole direction in the redshift ranges $0-0.2$, $0-0.4$,
 $0-0.6$ and $0-2.26$. The corresponding results are given in
 Table~\ref{tab2}. In all the four redshift ranges, no
 anisotropy has been found, since $A_D=0$ is fully consistent
 with the SNIa in each redshift range, and the $1\sigma$ region
 of $l$ and $b$ is the whole sky. This suggests that there is
 no (considerable) redshift anisotropy in the Pantheon sample.

As is mentioned in Sec.~\ref{sec1}, there are also anomalous
 anisotropies in the CMB map, which still persist in the Planck
 data~\cite{Ade:2013nlj,Ade:2015hxq,Schwarz:2015cma}. Since the
 CMB is extremely well sampled at all angles by the Planck
 satellite, these anomalous anisotropies cannot arise from lack
 of sky coverage as in the case of the SNIa samples. There are
 two possibilities that would render the CMB results compatible
 with the null results obtained in the present work, namely
 (1)~the CMB results are due to anisotropies at very
 high redshifts, such as an intrinsic dipole on the last
 scattering surface~\cite{Roldan:2016ayx}; (2)~the CMB results
 are due to anisotropies from structures at very low redshifts
 $z<0.03$ which do not give rise to a purely kinematic CMB
 dipole~\cite{Bolejko:2015gmk}. In fact, evidence for a
 non-kinematic dipole has been seen at the $99.5\%$ confidence
 level in number counts of radio galaxies~\cite{Rubart:2013tx},
 which is consistent with the second possibility and the
 result of~\cite{Bengaly:2017slg}.


 \begin{table}[tb]
 \renewcommand{\arraystretch}{1.5}
 \begin{center}
 \begin{tabular}{cccccc} \hline\hline

\ Redshift range\quad &  $N_{\rm SN}$  &  $\Omega_{m0}$  &  $A_D$  &   $l$ &   $b$ \  \\ \hline
\ $0-0.2$   &   411 &   $0.2493^{+0.0983}_{-0.1008}$    &    $  (0.0561^{+2.8835}_{-2.8926}) \times 10^{-3}  $ &   $  0^{\circ} \le  l  \le 360^{\circ}  $  &  $  -90^{\circ} \le  b \le 90^{\circ}  $  \\
\ $0-0.4$   &    766   &  $0.2968^{+0.0369}_{-0.0374}$   &  $  (0.0215^{+3.0505}_{-3.0803}) \times 10^{-3} $    &  $ 0^{\circ} \le  l \le 360^{\circ}$    &  $  -90^{\circ} \le b \le 90^{\circ} $     \\
\ $0-0.6$   &    886   &   $0.2865^{+0.0285}_{-0.0283}$   & $ (-0.0960^{+2.7986}_{-3.0973}) \times 10^{-3}  $   &  $   0^{\circ} \le l \le 360^{\circ} $   &   $   -90^{\circ} \le b \le 90^{\circ} $    \\
\ $0-2.26$  & \ \ \ 1048 \ \ \ & \ \ $0.2975^{+0.0219}_{-0.0218}$ \ \ & \ \ $(-0.0931^{+2.9619}_{-2.9316})\times 10^{-3}$ \ \  & \ \ $0^{\circ} \le l \le 360^{\circ}$ \ \  & \ $-90 ^{\circ} \le b \le 90^{\circ}$ \ \\

 \hline\hline
 \end{tabular}
 \end{center}
 \caption{\label{tab2} The redshift range, the number of SNIa,
 and the constraints on $\Omega_{m0}$, the dipole magnitude
 $A_D$, and the dipole direction $(l,\,b)$. See the text for
 details.}
 \end{table}



\section*{ACKNOWLEDGEMENTS}

We heartily thank the anonymous referee for all the very expert and
 useful comments and suggestions, which have significantly helped us
 to improve this work. We are grateful to D.~M.~Scolnic for
 private communication. We also thank Xiao-Bo~Zou, Zhao-Yu~Yin,
 Da-Chun~Qiang, Zhong-Xi~Yu, Shou-Long~Li, and Dong-Ze~Xue for
 kind help and discussions. This work was supported in part by
 NSFC under Grants No.~11575022 and No.~11175016.

\renewcommand{\baselinestretch}{1.0}



\begin{thebibliography}{99}

\bibitem{Weinberg}
S.~Weinberg, {\it Gravitation and Cosmology},
 John Wiley \& Sons, Inc., New York (1972);\\
S.~Weinberg, {\it Cosmology},
 Oxford University Press, Oxford (2008).

\bibitem{Kolb90}
E.~W.~Kolb and M.~S.~Turner,
 {\it The Early Universe}, Addison Wesley (1990).

\bibitem{Hogg:2004vw}
  D.~W.~Hogg {\it et al.},
  Astrophys.\ J.\  {\bf 624}, 54 (2005)
  [astro-ph/0411197].

\bibitem{Hajian:2006ud}
  A.~Hajian and T.~Souradeep,
  Phys.\ Rev.\ D {\bf 74}, 123521 (2006)
  [astro-ph/0607153];\\
  T.~R.~Jaffe {\it et al.},
  Astrophys.\ J.\  {\bf 629}, L1 (2005)
  [astro-ph/0503213].

\bibitem{Caldwell:2007yu}
  R.~R.~Caldwell and A.~Stebbins,
  Phys.\ Rev.\ Lett.\  {\bf 100}, 191302 (2008)
  [arXiv:0711.3459].

\bibitem{LTB}
  G.~Lema\^{\i}tre,
  Annales de la Soci\'et\'e Scientifique de Bruxelles
  A {\bf 53}, 51 (1933),
  see Gen.\ Rel.\ Grav.\  {\bf 29}, 641 (1997) for English translation;\\
  R.~C.~Tolman,
  Proc.\ Nat.\ Acad.\ Sci.\  {\bf 20}, 169 (1934),
  see Gen.\ Rel.\ Grav.\  {\bf 29}, 935 (1997) for English translation;\\
  H.~Bondi,
  Mon.\ Not.\ Roy.\ Astron.\ Soc.\  {\bf 107}, 410 (1947).

\bibitem{Yan:2014eca}
  X.~P.~Yan, D.~Z.~Liu and H.~Wei,
  Phys.\ Lett.\ B {\bf 742}, 149 (2015)
  [arXiv:1411.6218].

\bibitem{Deng:2018yhb}
  H.~K.~Deng and H.~Wei,
  Phys.\ Rev.\ D {\bf 97}, no. 12, 123515 (2018)
  [arXiv:1804.03087].

\bibitem{Godel:1949ga}
  K.~G\"odel,
  Rev.\ Mod.\ Phys.\  {\bf 21}, 447 (1949).

\bibitem{Bianchi}
  S.~Kumar and C.~P.~Singh,
  Astrophys.\ Space Sci.\  {\bf 312}, 57 (2007);\\
  R.~Venkateswarlu and K.~Sreenivas,
  Int.\ J.\ Theor.\ Phys.\  {\bf 53}, 2051 (2014);\\
  S.~Ram and C.~P.~Singh,
  Astrophys.\ Space Sci.\  {\bf 257}, 287 (1998);\\
  L.~Yadav, V.~K.~Yadav and T.~Singh,
  Int.\ J.\ Theor.\ Phys.\  {\bf 51}, 3113 (2012);\\
  A.~Pradhan and H.~Amirhashchi,
  Astrophys.\ Space Sci.\  {\bf 332}, 441 (2011)
  [arXiv:1010.2362];\\
  D.~K.~Banik, S.~K.~Banik and K.~Bhuyan,
  Astrophys.\ Space Sci.\  {\bf 362}, 51 (2017);\\
  B.~Mishra, P.~K.~Sahoo and S.~Suresh,
  Astrophys.\ Space Sci.\  {\bf 358}, 7 (2015);\\
  A.~K.~Yadav,
  Astrophys.\ Space Sci.\  {\bf 335}, 565 (2011)
  [arXiv:1101.4349];\\
  B.~Saha,
  Int.\ J.\ Theor.\ Phys.\  {\bf 52}, 3646 (2013)
  [arXiv:1209.6029];\\
  J.~M.~Bradley and E.~Sviestins,
  Gen.\ Rel.\ Grav.\  {\bf 16}, 1119 (1984);\\
  D.~Lorenz,
  Phys.\ Rev.\ D {\bf 22}, 1848 (1980);\\
  D.~Sofuoglu,
  Astrophys.\ Space Sci.\  {\bf 361}, 12 (2016);\\
  B.~Mishra and S.~K.~Tripathy,
  Mod.\ Phys.\ Lett.\ A {\bf 30}, no. 36, 1550175 (2015)
  [arXiv:1507.03515];\\
  B.~Mishra {\it et al.},
  Adv.\ High Energy Phys.\  {\bf 2018}, 6306848 (2018)
  [arXiv:1706.07661];\\
  B.~Mishra, S.~K.~Tripathy and P.~P.~Ray,
  Astrophys.\ Space Sci.\  {\bf 363}, 86 (2018)
  [arXiv:1701.08632].

\bibitem{Li:2015uda}
  X.~Li, H.~N.~Lin, S.~Wang and Z.~Chang,
  Eur.\ Phys.\ J.\ C {\bf 75}, no. 5, 181 (2015)
  [arXiv:1501.06738];\\
  Z.~Chang, S.~Wang and X.~Li,
  Eur.\ Phys.\ J.\ C {\bf 72}, 1838 (2012)
  [arXiv:1106.2726].

\bibitem{Axisofevil}
  K.~Land and J.~Magueijo,
  Phys.\ Rev.\ Lett.\  {\bf 95}, 071301 (2005) [astro-ph/0502237];\\
  K.~Land and J.~Magueijo,
  Mon.\ Not.\ Roy.\ Astron.\ Soc.\  {\bf 357}, 994 (2005) [astro-ph/0405519];\\
  K.~Land and J.~Magueijo,
  Mon.\ Not.\ Roy.\ Astron.\ Soc.\  {\bf 378}, 153 (2007)
  [astro-ph/0611518].

\bibitem{Zhao:2016fas}
  W.~Zhao and L.~Santos,
  The Universe, no. 3, 9 (2015)
  [arXiv:1604.05484].

\bibitem{Hansen:2004vq}
  F.~K.~Hansen {\it et al.},
  Mon.\ Not.\ Roy.\ Astron.\ Soc.\  {\bf 354}, 641 (2004)
  [astro-ph/0404206].

\bibitem{Schwarz:2007wf}
  D.~J.~Schwarz and B.~Weinhorst,
  Astron.\ Astrophys.\  {\bf 474}, 717 (2007)
  [arXiv:0706.0165].

\bibitem{Antoniou:2010gw}
  I.~Antoniou and L.~Perivolaropoulos,
  JCAP {\bf 1012}, 012 (2010)
  [arXiv:1007.4347].

\bibitem{Mariano:2012wx}
  A.~Mariano and L.~Perivolaropoulos,
  Phys.\ Rev.\ D {\bf 86}, 083517 (2012)
  [arXiv:1206.4055].

\bibitem{Cai:2011xs}
  R.~G.~Cai and Z.~L.~Tuo,
  JCAP {\bf 1202}, 004 (2012)
  [arXiv:1109.0941];\\
  R.~G.~Cai, Y.~Z.~Ma, B.~Tang and Z.~L.~Tuo,
  Phys.\ Rev.\ D {\bf 87}, 123522 (2013)
  [arXiv:1303.0961].

\bibitem{Zhao:2013yaa}
  W.~Zhao, P.~X.~Wu and Y.~Zhang,
  Int.\ J.\ Mod.\ Phys.\ D {\bf 22}, 1350060 (2013)
  [arXiv:1305.2701].

\bibitem{Yang:2013gea}
  X.~Yang, F.~Y.~Wang and Z.~Chu,
  Mon.\ Not.\ Roy.\ Astron.\ Soc.\  {\bf 437}, 1840 (2014)
  [arXiv:1310.5211].

\bibitem{Chang:2014nca}
  Z.~Chang and H.~N.~Lin,
  Mon.\ Not.\ Roy.\ Astron.\ Soc.\  {\bf 446}, 2952 (2015)
  [arXiv:1411.1466].

\bibitem{Lin:2015rza}
  H.~N.~Lin, S.~Wang, Z.~Chang and X.~Li,
  Mon.\ Not.\ Roy.\ Astron.\ Soc.\  {\bf 456}, 1881 (2016)
  [arXiv:1504.03428].

\bibitem{Lin:2016jqp}
  H.~N.~Lin, X.~Li and Z.~Chang,
  Mon.\ Not.\ Roy.\ Astron.\ Soc.\  {\bf 460}, no. 1, 617 (2016)
  [arXiv:1604.07505].

\bibitem{Chang:2017bbi}
  Z.~Chang, H.~N.~Lin, Y.~Sang and S.~Wang,
  arXiv:1711.11321 [astro-ph.CO].

\bibitem{Javanmardi:2015sfa}
  B.~Javanmardi {\it et al.},
  Astrophys.\ J.\  {\bf 810}, no. 1, 47 (2015)
  [arXiv:1507.07560].

\bibitem{Bengaly:2015dza}
  C.~A.~P.~Bengaly, A.~Bernui and J.~S.~Alcaniz,
  Astrophys.\ J.\  {\bf 808}, 39 (2015)
  [arXiv:1503.01413];\\
  U.~Andrade {\it et al.},
  Phys.\ Rev.\ D {\bf 97}, no. 8, 083518 (2018)
  [arXiv:1711.10536].

\bibitem{Wang:2017ezt}
  Y.~Y.~Wang and F.~Y.~Wang,
  Mon.\ Not.\ Roy.\ Astron.\ Soc.\  {\bf 474}, no. 3, 3516 (2018)
  [arXiv:1711.05974].

\bibitem{Sun:2018epo}
  Z.~Q.~Sun and F.~Y.~Wang,
  arXiv:1804.05191 [astro-ph.CO].

\bibitem{Meszaros:2009ux}
  A.~Meszaros {\it et al.},
  AIP Conf.\ Proc.\  {\bf 1133}, 483 (2009)
  [arXiv:0906.4034].

\bibitem{Wang:2014vqa}
  J.~S.~Wang and F.~Y.~Wang,
  Mon.\ Not.\ Roy.\ Astron.\ Soc.\  {\bf 443}, no. 2, 1680 (2014)
  [arXiv:1406.6448].

\bibitem{Chang:2014jza}
  Z.~Chang, X.~Li, H.~N.~Lin and S.~Wang,
  Mod.\ Phys.\ Lett.\ A {\bf 29}, 1450067 (2014)
  [arXiv:1405.3074].

\bibitem{Webb98}
  J.~K.~Webb {\it et al.},
  Phys.\ Rev.\ Lett.\  {\bf 82}, 884 (1999) [astro-ph/9803165].

\bibitem{Webb00}
  J.~K.~Webb {\it et al.},
  Phys.\ Rev.\ Lett.\  {\bf 87}, 091301 (2001) [astro-ph/0012539];\\
  M.~T.~Murphy {\it et al.},
  Mon.\ Not.\ Roy.\ Astron.\ Soc.\  {\bf 327}, 1208 (2001)
  [astro-ph/0012419].

\bibitem{Uzan10}
  J.~P.~Uzan,
  Living Rev.\ Rel.\  {\bf 14}, 2 (2011) [arXiv:1009.5514].

\bibitem{Barrow09}
  J.~D.~Barrow,
  Ann.\ Phys.\  {\bf 19}, 202 (2010)
  [arXiv:0912.5510].

\bibitem{HWalpha}
  H.~Wei,
  Phys.\ Lett.\ B {\bf 682}, 98 (2009) [arXiv:0907.2749];\\
  H.~Wei, X.~P.~Ma and H.~Y.~Qi,
  Phys.\ Lett.\ B {\bf 703}, 74 (2011) [arXiv:1106.0102].

\bibitem{Wei:2016moy}
  H.~Wei, X.~B.~Zou, H.~Y.~Li and D.~Z.~Xue,
  Eur.\ Phys.\ J.\ C {\bf 77}, no. 1, 14 (2017)
  [arXiv:1605.04571];\\
  H.~Wei and D.~Z.~Xue,
  Commun.\ Theor.\ Phys.\  {\bf 68}, no. 5, 632 (2017)
  [arXiv:1706.04063].

\bibitem{King:2012id}
  J.~A.~King {\it et al.},
  Mon.\ Not.\ Roy.\ Astron.\ Soc.\  {\bf 422}, 3370 (2012)
  [arXiv:1202.4758].

\bibitem{Webb:2010hc}
  J.~K.~Webb {\it et al.},
  Phys.\ Rev.\ Lett.\  {\bf 107}, 191101 (2011)
  [arXiv:1008.3907].

\bibitem{Zhou:2017lwy}
  Y.~Zhou, Z.~C.~Zhao and Z.~Chang,
  Astrophys.\ J.\  {\bf 847}, no. 2, 86 (2017)
  [arXiv:1707.00417].

\bibitem{Chang:2018vxs}
  Z.~Chang, H.~N.~Lin, Z.~C.~Zhao and Y.~Zhou,
  arXiv:1803.08344 [astro-ph.CO].

\bibitem{Singal:2013aga}
  A.~K.~Singal,
  Astrophys.\ Space Sci.\  {\bf 357}, no. 2, 152 (2015)
  [arXiv:1305.4134].

\bibitem{Bengaly:2017slg}
  C.~A.~P.~Bengaly, R.~Maartens and M.~G.~Santos,
  JCAP {\bf 1804}, 031 (2018)
  [arXiv:1710.08804].

\bibitem{Hutsemekers}
  D.~Hutsemekers {\it et al.},
  Astron.\ Astrophys.\  {\bf 441}, 915 (2005)
  [astro-ph/0507274];\\
  D.~Hutsemekers and H.~Lamy,
  Astron.\ Astrophys.\  {\bf 367}, 381 (2001)
  [astro-ph/0012182];\\
  D.~Hutsemekers {\it et al.},
  ASP Conf.\ Ser.\  {\bf 449}, 441 (2011) [arXiv:0809.3088]

\bibitem{Pelgrims:2016mhx}
  V.~Pelgrims,
  arXiv:1604.05141 [astro-ph.CO].

\bibitem{Amanullah:2010vv}
  R.~Amanullah {\it et al.},
  Astrophys.\ J.\  {\bf 716}, 712 (2010)
  [arXiv:1004.1711].

\bibitem{Suzuki:2011hu}
  N.~Suzuki {\it et al.},
  Astrophys.\ J.\  {\bf 746}, 85 (2012)
  [arXiv:1105.3470].

\bibitem{Betoule:2014frx}
  M.~Betoule {\it et al.},
  Astron.\ Astrophys.\  {\bf 568}, A22 (2014)
  [arXiv:1401.4064].

\bibitem{Wei:2010wu}
  H.~Wei,
  JCAP {\bf 1008}, 020 (2010)
  [arXiv:1004.4951].

\bibitem{Liu:2014vda}
  J.~Liu and H.~Wei,
  Gen.\ Rel.\ Grav.\  {\bf 47}, no. 11, 141 (2015)
  [arXiv:1410.3960].

\bibitem{Lelli:2016zqa}
  F.~Lelli, S.~S.~McGaugh and J.~M.~Schombert,
  Astron.\ J.\  {\bf 152}, 157 (2016)
  [arXiv:1606.09251].

\bibitem{Scolnic:2017caz}
  D.~M.~Scolnic {\it et al.},
  Astrophys.\ J.\  {\bf 859}, no. 2, 101 (2018)
  [arXiv:1710.00845].

\bibitem{Pantheondata}
The numerical data of the full Pantheon SNIa sample are available at\\
  http:$/\!/$dx.doi.org/10.17909/T95Q4X \\
  https:$/\!/$archive.stsci.edu/prepds/ps1cosmo/index.html

\bibitem{Pantheonplugin}
The Pantheon plugin for CosmoMC is available at\\
  https:$/\!/$github.com/dscolnic/Pantheon

\bibitem{Conley:2011ku}
  A.~Conley {\it et al.},
  Astrophys.\ J.\ Suppl.\  {\bf 192}, 1 (2011)
  [arXiv:1104.1443].

\bibitem{Wang:2015tua}
  Y.~Wang and M.~Dai,
  Phys.\ Rev.\ D {\bf 94}, no. 8, 083521 (2016)
  [arXiv:1509.02198].

\bibitem{Zou:2017ksd}
  X.~B.~Zou, H.~K.~Deng, Z.~Y.~Yin and H.~Wei,
  Phys.\ Lett.\ B {\bf 776}, 284 (2018)
  [arXiv:1707.06367].

\bibitem{Panradec}
The position of each SNIa in the Pantheon sample can be found
 in~\cite{Pantheondata} or \cite{Pantheonplugin}. In particular, the
 right ascension (ra) and declination (dec)
 in degree (equatorial coordinate system, J2000) of the first
 1030 SNIa are available at~\cite{Pantheondata}\ \
 https:$/\!/$archive.stsci.edu/hlsps/ps1cosmo/scolnic/data$_{-}$fitres
 \ or the folder {\tt data$_{-}$fitres} in the Pantheon
 plugin~\cite{Pantheonplugin}. The ra and dec of the last 18 SNIa can
 be found in the other databases, such as the Open Supernova
 Catalog at\ \ https:$/\!/$sne.space

\bibitem{Lewis:2002ah}
  A.~Lewis and S.~Bridle,
  Phys.\ Rev.\ D {\bf 66}, 103511 (2002)
  [astro-ph/0205436];\\
  http:$/\!/$cosmologist.info/cosmomc

\bibitem{Gorski:2004by}
  K.~M.~Gorski {\it et al.},
  Astrophys.\ J.\  {\bf 622}, 759 (2005)
  [astro-ph/0409513];\\
  http:$/\!/$healpix.sourceforge.net

\bibitem{Multipole}
https:$/\!/$en.wikipedia.org/wiki/Multipole$_{-}$expansion

\bibitem{Thompson:1994}
W.~J.~Thompson, {\it Angular Momentum}, Wiley-VCH (1994).

\bibitem{Sun:2018cha}
  Z.~Q.~Sun and F.~Y.~Wang,
  the first version (v1) of arXiv:1805.09195 [astro-ph.CO].\\
In fact, after we pointed out their mistakes, they have accordingly
 modified their later arXiv versions and the published journal version. So,
 ``SW18'' means intendedly the first version (v1) of arXiv:1805.09195.

\bibitem{disSW18}
Note that our Fig.~\ref{fig1} is quite different from Fig.~1 of
 SW18~\cite{Sun:2018cha}. We can check it as follows. It is easy to
 verify that most of the JLA SNIa~\cite{Betoule:2014frx} are
 also included in the Pantheon sample. The distribution of
 740 JLA SNIa in the galactic coordinate system was shown in
 Fig.~1 of~\cite{Wang:2017ezt}, which is very similar to our
 Fig.~\ref{fig1} but is quite different from Fig.~1 of
 SW18~\cite{Sun:2018cha}. Since the difference between the JLA
 sample and the Pantheon sample is not so significant, we speculate
 that something might be wrong in Fig.~1 of SW18~\cite{Sun:2018cha}.

\bibitem{Heneka:2013hka}
  C.~Heneka, V.~Marra and L.~Amendola,
  Mon.\ Not.\ Roy.\ Astron.\ Soc.\  {\bf 439}, 1855 (2014)
  [arXiv:1310.8435].

\bibitem{Campanelli:2006vb}
  L.~Campanelli, P.~Cea and L.~Tedesco,
  Phys.\ Rev.\ Lett.\  {\bf 97}, 131302 (2006)
  [astro-ph/0606266];\\
  L.~Campanelli, P.~Cea and L.~Tedesco,
  Phys.\ Rev.\ D {\bf 76}, 063007 (2007)
  [arXiv:0706.3802].

\bibitem{bbc}
  D.~Scolnic and R.~Kessler,
  Astrophys.\ J.\  {\bf 822}, no. 2, L35 (2016)
  [arXiv:1603.01559];\\
  R.~Kessler and D.~Scolnic,
  Astrophys.\ J.\  {\bf 836}, no. 1, 56 (2017)
  [arXiv:1610.04677].

\bibitem{Andrade:2018eta}
  U.~Andrade, C.~A.~P.~Bengaly, B.~Santos and J.~S.~Alcaniz,
  arXiv:1806.06990 [astro-ph.CO].

\bibitem{Ade:2013nlj}
  P.~A.~R.~Ade {\it et al.},
  Astron.\ Astrophys.\  {\bf 571}, A23 (2014)
  [arXiv:1303.5083].

\bibitem{Ade:2015hxq}
  P.~A.~R.~Ade {\it et al.},
  Astron.\ Astrophys.\  {\bf 594}, A16 (2016)
  [arXiv:1506.07135].

\bibitem{Schwarz:2015cma}
  D.~J.~Schwarz {\it et al.},
  Class.\ Quant.\ Grav.\  {\bf 33}, no. 18, 184001 (2016)
  [arXiv:1510.07929].

\bibitem{Roldan:2016ayx}
  O.~Roldan, A.~Notari and M.~Quartin,
  JCAP {\bf 1606}, no. 06, 026 (2016)
  [arXiv:1603.02664].

\bibitem{Bolejko:2015gmk}
  K.~Bolejko, M.~A.~Nazer and D.~L.~Wiltshire,
  JCAP {\bf 1606}, no. 06, 035 (2016)
  [arXiv:1512.07364].

\bibitem{Rubart:2013tx}
  M.~Rubart and D.~J.~Schwarz,
  Astron.\ Astrophys.\  {\bf 555}, A117 (2013)
  [arXiv:1301.5559].

\bibitem{revfoot}
We use the data file {\tt lcparam$_{-}$full$_{-}$long.txt} in
 the Pantheon plugin~\cite{Pantheonplugin}, equivalent to the file\\
 {\tt hlsp$_{-}$ps1cosmo$_{-}$panstarrs$_{-}$gpc1$_{-}$all$_{-}$model$_{-}$v1$_{-}$lcparam-full.txt}
 in the Pantheon data repository~\cite{Pantheondata}, which corresponds to
 Table~17 of~\cite{Scolnic:2017caz}.

\end{thebibliography}
\end{document}